\documentclass[preprint,aps,12pt]{revtex4}
\usepackage{amsmath}
\usepackage{amssymb}
\usepackage{epsfig}
\usepackage{slashed}
\usepackage{graphicx}
\usepackage{multirow,color}
\graphicspath{{figure/}}
\def\missE{\slashed E_T} 
\begin{document}
\title{ The signatures of doubly charged leptons in future linear colliders}
\author{Yu-Chen Guo}\email{lgguoyuchen@126.com}
\author{Chong-Xing Yue}\email{cxyue@lnnu.edu.cn}
\author{Zhi-Cheng Liu}

\affiliation{
Department of Physics, Liaoning Normal University, Dalian 116029,  China
\vspace*{1.5cm}}

\begin{abstract}
We discuss the production of the doubly charged leptons in future linear electron positron colliders, such as the International Linear Collider and Compact Linear Collider. Such states are introduced in extended weak-isospin multiplets by composite models. We discuss the production cross section of $e^{-}\gamma\rightarrow L^{--}W^{+}$ and carry out analyses for hadronic, semi-leptonic and pure leptonic channels based on the full simulation performance of the Silicon Detector. The 3- and 5-sigma statistical significance exclusion curves are provided in the model parameter space. It is found that the hadronic channel could offer the most possible detectable signature.

\end{abstract}


\maketitle
\section{Introduction}

Although the standard model (SM) can successfully describe many phenomena up to the energies that can be reached today, there are sufficient reasons to believe that the SM is an effective theory. Particle dark matter, neutrino masses, and the Higgs mechanism operating at a scale much lower than the Planck scale are among the pressing issues.
Especially whether the proliferation of fermionic states are fundamental particles has been one of the motivations for many theoretical considerations beyond the standard model. A natural explanation for the replication of fermionic generations could be that they are not truly fundamental particles but instead bound states of some unknown constituents.
Compositeness of ordinary fermions has been investigated phenomenologically for quite long time \cite{history1,history2,history3,history4,history5}.
It predicts that excited states arise as a set of composites of some more fundamental particles, and they are heavier than the known fermions. Any signal for such kinds of new fermions will play an important role in testing the SM flavor structure and discovering new physics.

Most of the excited fermions searches employ the theoretical framework of gauge mediated effective Lagrangians, concentrating on multiplets of weak isospin $I_{W}=0$ and $I_{W}=1/2$. Considering another aspect of compositeness, the weak isospin invariance, the usual singlet ($I_{W}=0$) and doublet ($I_{W}=1/2$) isospin values are extended to include $I_{W}=1$ and $I_{W}=3/2$ by the extended isospin model \cite{history4}. Perhaps the most interesting consequence of this model is the existence of the heavy charged leptons, which are expected to produce possible signatures. Since such excited states are vector-like and do not couple to the Higgs at tree level, their electroweak charge causes no conflict with the measurement of the $S$ and $T$ parameters \cite{oblique1,oblique2}. Before $SU(2)\otimes U(1)$ breaking these excited fermions are allowed to acquire their masses, which avoids the dangerous bounds coming from the precise determinations of $\Delta\rho$.


Up to now, all direct searches within the reach of the experiments have been carried out, as yet, with no evidence of excited leptons. Signatures of excited leptons have been searched for at the DESY Hadron Electron Ring Accelerator (HERA) \cite{DESY1,DESY2,DESY3}, the CERN Large Electron Positron collider (LEP) \cite{lep1,lep2,lep3,lep4} and the Fermilab Tevatron collider \cite{fermi1,fermi2}. As a result, the bounds of $\mathcal{O}(200)$ GeV (by HERA and LEP) and 430 GeV for the gauge mediated model (by Tevatron) are set on the excited lepton masses at the $95\%$ confidence level.
The possibility to search excited quarks and leptons at the Tevatron and the LHC is discussed in \cite{1984,1990}, where the authors estimated the production rates of various signatures. The possible signals of the excited fermions with $I_{W}=0$ and $I_{W}=1/2$ have been analyzed at compositeness scales of order $1-2$ TeV in \cite{2002}. With the same isospin, in the quark sector, the LHC experiments \cite{atlas1,cms1} have put excluding bounds on the mass of excited quarks up to 2 TeV, and in the lepton sector, the generic searches for leptonic final states at 8 TeV LHC \cite{cms3} exclude the masses of the exotic leptons up to 2.45 TeV. Such searches strongly depend on the flavor structure. Light states are still allowed if their couplings are suppressed, or they decay into final states affected by large backgrounds, or they are not efficiently produced at the LHC.
The ATLAS and CMS collaborations have searched for long-lived doubly charged states by Drell-Yan-like pair production. From the ATLAS, the long-lived doubly charged state masses have been excluded up to 430 GeV based on the run at $\sqrt{s}=7$ TeV with $L=5$ fb$^{-1}$\cite{atlas4}. Similarly the CMS Collaboration sets lower mass limits up to 685 GeV within the run at $\sqrt{s}=8$ TeV with $L=18.8$ fb$^{-1} $\cite{cms4}. Nevertheless, these limits do not hold for promptly decaying doubly charged particles.


Phenomenology of doubly charged leptons at LHC has been carried out in a number of publications \cite{PRD42:815,1201.3764,1405.3911,CPL.DCL.LHC,1306.2066,1105.6299,HanTao,1307.1711,ma1,ma2,PRD91.Yue,1508.01046,1602.07519,1609.05251,1610.06587,EPJC76:593}. As a hadron collider, LHC has limitations to undertake precision studies. Production of the doubly charged leptons in linear electron-positron and electron-proton colliders, which offers much cleaner experimental collisions, was addressed already long ago \cite{linearhistory1,linearhistory2}. A future linear electron positron collider, such as the International Linear Collider (ILC) \cite{ILC} and the Compact Linear Collider (CLIC) \cite{CLIC1,CLIC2}, is attracting growing attention \cite{1411.6556,DCL.ee.Yue,PRD77}. Along with an $e^{+}e^{-}$ collider, other options such as $e^{-}e^{-}$, $e^{-}\gamma$ and $\gamma\gamma$ colliders have also been discussed \cite{er,rPDF,PRD65:033009}. The $e^-\gamma$ option is based on $e^{+}e^{-}$ collisions, where one of the electron beams is converted to the photon beam. We are going to investigate associated production of a single doubly charged lepton with a $W$ boson via $e^{-}\gamma$ collisions, and discuss the most promising signature for hadronic, semi-leptonic and pure leptonic channels. To make our analysis more realistic in future liner colliders, it would be desirable to introduce the effects of detectors. The Silicon Detector (SiD) is one of the two detectors described in the ILC Technical Design Report (TDR) \cite{ILCTDR}. The new generation detector at the CLIC is expected to perform at least at the same level as the SiD of the ILC. In this paper, our analysis bases on the DSiD \cite{sid}, a fast simulation Delphes detector for the ILC based on the full simulation performance of the SiD.

The rest of the paper is organized as follows: in Sect. II, we review the main features of the extended isospin model; in Sect. III, we calculate the cross sections of $L^{--}W^{+}$ production in future linear colliders; in Sect. IV,  we present the results of the fast simulation for three channels and discuss the sensitivity to the model parameters for the best channel in detail; finally Sect. V summarizes the analysis results and gives our conclusions.

\section{The main features of the extended isospin model}
The details of the extended isospin model have been studied in the original work Ref.~\cite{model}, including the particle spectrum, all couplings and interactions. Here we will briefly review the essential features of the model and focus our attention on the main features of the higher multiplets ($I_{W}=1,3/2$).

Before we learnt about quarks and gluons, the isospin was generally used to discuss the possible patterns of baryon and meson resonances. Similarly, weak isospin can reveal some properties of hypothetical excited fermions whatever the internal structure is.
First of all, we need to know the allowed isospin states. The standard model fermions have $I_{W}=0$ and $I_{W}=1/2$, and the electroweak bosons have $I_{W}=0$ and $I_{W}=1$. By combining them, fermionic excited states are extended to $I_{W}=1$ and $I_{W}=3/2$, which include the doubly charged leptons. The doubly charged leptons belong to the following isospin multiplets,
\begin{eqnarray}
L_1 = \left( \begin{array}{l}
L^{0} \\
L^{-} \\
L^{--} \end{array} \right) ,
\qquad
L_{3/2} = \left( \begin{array}{l}
L^{+} \\
L^{0} \\
L^{-} \\
L^{--} \end{array} \right),
\end{eqnarray}
with similar multiplets for the antiparticles. These higher multiplets can couple to the known light fermions only through the $W$ boson, because they contribute to the isovector current rather than the hypercharge current. In order to describe couplings between the doubly charged lepton and SM leptons, we use the gauge-mediated (GM) model Lagrangians which are made of dimension five operators with the compositeness scale $\Lambda$,
\begin{align}
\label{lag1}
\mathcal{L}_{\text{G}}^{(1)}=i\frac{g\, f}{\Lambda}\left( \bar{L} \, \sigma_{\mu \nu} \, \partial^{\nu} \, W^{\mu} \,  \frac{1+\gamma^{5}}{2} \, \ell \right)  + h. c.\\
\mathcal{L}_{\text{G}}^{(3/2)}=
 i\frac{g\,\tilde{f}}{\Lambda}\left( \bar{L}\sigma_{\mu \nu} \, \partial^{\nu} \, W^{\mu} \, \frac{1-\gamma^{5}}{2} \, \ell  \right)  + h.c.,
 \end{align}
where $f$, $\tilde{f}$ are coupling constants that parametrize the effective interaction for the $I_{W}=1$ and $I_{W}=3/2$, $g$ is the SU(2) coupling, and $\sigma^{\mu \nu}=i\left[\gamma^{\mu},\gamma^{\nu} \right] /2$. $f$ and $\tilde{f}$ are usually set to one and we keep this choice as well. From the GM Lagrangians, the only available decay channel of the doubly charged lepton is $L^{--}\rightarrow W^-\ell^-$.

It is well known that contact interactions (CIs) play a major role in the production and decay of the excited states. These can be introduced \cite{1990} by an effective dimension 6 four-fermion Lagrangian of the type:
\begin{align}
\label{lag2}
\mathcal{L}_{\text{CI}}=\frac{g_*^2}{\Lambda^2}\frac{1}{2}j^\mu j_\mu,
\end{align}
where the current reads
\begin{align}
\label{j1}
j_\mu=\eta_L\overline{f}_L\gamma_\mu f_L+\eta'_{L}\overline{f}^*_L\gamma_\mu f^*_L+\eta''_{L}\overline{f}^{*}_{L}\gamma_\mu f_L+h.c.+(L\rightarrow R),
\end{align}
where $g^2_*$ is conventionally chosen equal to $4\pi$ and the $\eta$ factors are usually set equal to one. The $f_L$ stands for a SM fermion and $f^*_L$ for an excited fermion. We neglect the right-handed currents for simplicity. For the decay of the bi-lepton and the diquark, we need the following current:
\begin{align}
\label{j2}
j_\mu=\overline{\nu}_L\gamma_\mu \ell_L+\overline{q}_L\gamma_\mu q_L+\overline{L}_L\gamma_\mu \ell_L+h.c.
\end{align}
The corresponding Lagrangian reads
\begin{align}
\label{lag3}
\mathcal{L}_{\text{CI}}=\frac{g_*^2}{\Lambda^2}(\overline{\nu}_L\gamma_\mu \ell_L\overline{L}_L\gamma_\mu \ell_L+\overline{q}_L\gamma_\mu q_L\overline{L}_L\gamma_\mu \ell_L).
\end{align}

\section{Production of the doubly charged leptons in future linear colliders}
Projects for future linear $e^+e^-$ colliders include the possibility of $e\gamma$ and $\gamma\gamma$ options. This idea was put forward by the author in Ref. \cite{er}. The required high-energy photons were suggested to be obtained by backward Compton scattering of laser light on the electron beams of colliders.

The cross sections $\sigma(s)$ can be evaluated from $\hat{\sigma}(\hat{s})$ by convoluting with the backscattered laser photon spectrum $f_{\gamma/e}(x)$:
\begin{eqnarray}
\sigma(s)=\int \mathrm{d}x f_{\gamma/e}(x,\frac{\sqrt{s}}{2}) \hat{\sigma}(\hat{s}).
\end{eqnarray}
The backscattered photon distribution function is given in Ref. \cite{rPDF}. In that case the photon beam carries on average a fraction of $80\%$ of the electron beam, but with a certain width, which leads to a maximum $e\gamma$ center of mass energy of $\sim0.91\times\sqrt{s}$.

\begin{figure}[htb]
\includegraphics [scale=1] {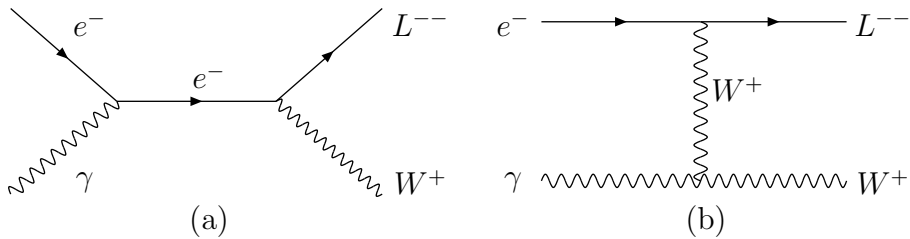}
\caption{ Feynman diagrams for the process $e^{-}\gamma\rightarrow L^{--}W^{+}$. }
\label{Feynman}
\end{figure}

Doubly charged leptons are the most characteristic particles in the extended isospin model. As shown in Fig. 1, they can be associated produced with a $W$ boson by the $s-$ and $t-$channels via $e^{-}\gamma$ collisions. The process $e^{-}\gamma\rightarrow L^{--}W^{+}$ is better than pair production by $e^+e^-$ collisions because of a wider mass parameter space available to probe.

In order to perform the needed numerical calculations, we have used FeynRules \cite{Feynrules} to generate the Feynman rules of the effective Lagrangian. The cross sections have been calculated using the CalcHEP \cite{calchep} generator.
We present the production cross sections versus the doubly charged lepton mass $M_{L}$ at the 1 TeV ILC and 3 TeV CLIC in Fig. 2. The plots show that the cross section of $L^{--}W^{+}$ production decreases with the increase of the doubly charged lepton mass $M_{L}$. It is worth mentioning that the results are the same for both isospin values $I_W=1$ and $I_W=3/2$, due to the structure of partonic cross section. The bound by the current experiments brings us to a quite narrow window for discussing the masses of doubly charged leptons at the ILC. However, even if the invariant masses of new particles exceeds the center of mass energy of the ILC, the virtual doubly charged leptons could also offer a possibility to produce signatures for indirect searches in future colliders.
\begin{figure}[!htb]
\begin{center}
\includegraphics [scale=0.82] {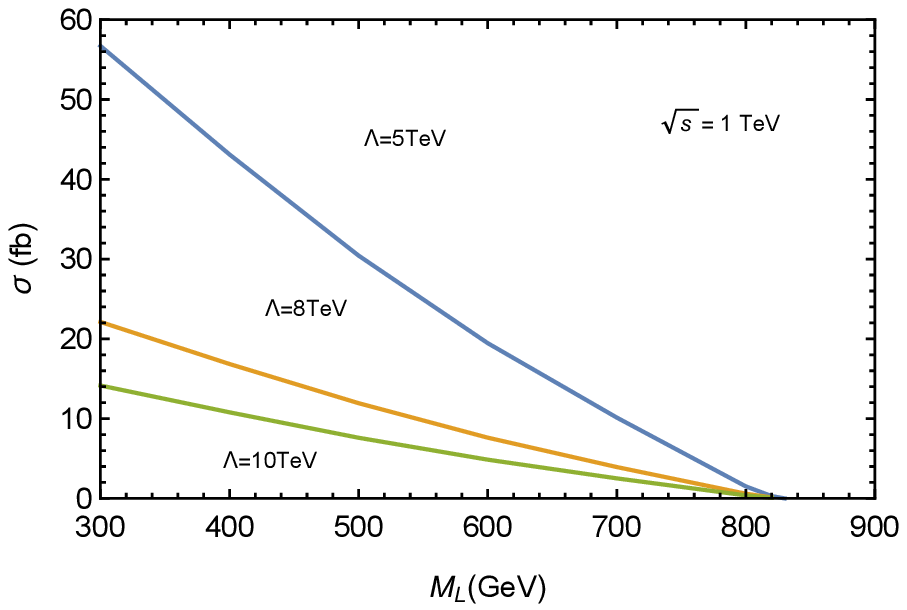}
\includegraphics [scale=0.82] {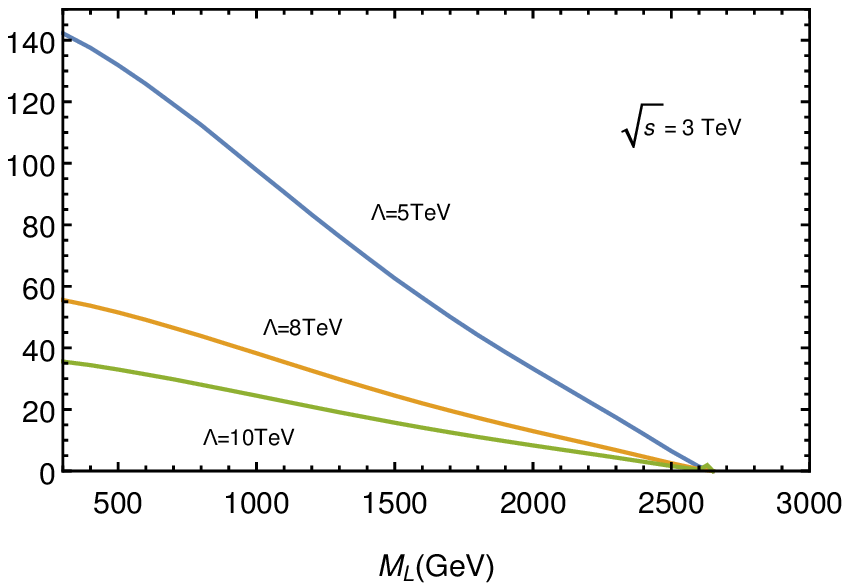}
\caption{The cross sections of $e^{-}\gamma\rightarrow L^{--} W^{+}$ as a function of the doubly charged lepton mass for different values of the compositeness scale $\Lambda$. We set the center of mass energy is 1 TeV for the ILC (left) and 3 TeV for the CLIC (right). } \label{ILC}
\end{center}
\label{fig2}
\end{figure}

\section{Signal and background simulation}
\begin{figure}[htb]
\includegraphics [scale=0.8] {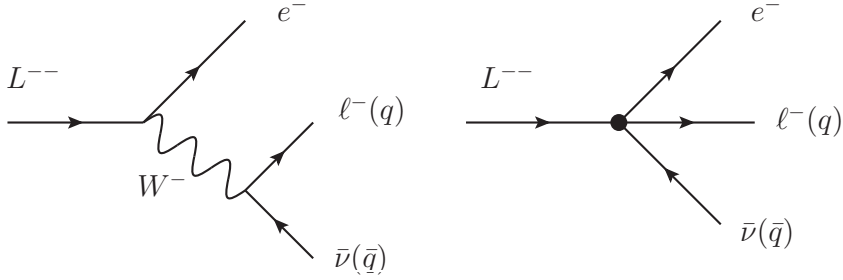}
\caption{ The Feynman diagrams for the doubly charged lepton decay process induced by gauge (left) and contact (right) interactions.}
\label{Feynman2}
\end{figure}

In this section we discuss the signals of doubly charged lepton production and the corresponding backgrounds in future colliders. The four fermion contact interactions could contribute to the process $e^{-} \gamma \rightarrow L^{--} W^{+}$ via the triangle loop diagrams, which are neglected in this work because of the loop suppression. However, contact interactions will play a major role in the decay of the doubly charged leptons \cite{1984,1990,linearhistory2,1201.3764,EPJC76:593}. The signatures can be realised by two kinds of Feynman diagrams via both gauge and contact interactions which are shown in Fig. \ref{Feynman2}.
To obtain results as general as possible, all the combinations of contact and gauge driven decay processes are therefore considered together with their interferences.

The final state signature generated by the decays of doubly charged lepton and $W$ boson. Because of the multiple decay modes, we classify the signals in terms of the charged lepton multiplicity. In the following, we will carry out analyses for hadronic, semi-leptonic and pure leptonic channels based on the fast simulation performance of SiD, focusing on two typical cases of $M_L=500$ and 700 GeV with $\Lambda=5$ TeV at the 1 TeV ILC and 3 TeV CLIC. Fast detector simulations are performed by PYTHIA8 \cite{py8} and Delphes3 \cite{Delphes} with the DSiD detector card \cite{sid}. Jets are clustered by using the anti-$k_{t}$ algorithm with a cone radius $\Delta R = 0.7$.

\subsection{Hadronic Channel}

In the hadronic decay channel, production of the doubly charged lepton in association with a $W$ boson can provide a distinct signal $e^-+4j$,
\begin{eqnarray}
e^{-}\gamma\rightarrow L^{--}W^{+}\rightarrow e^{-}jjjj.
\end{eqnarray}
The hadronic channel is the main decay channel of $L^{--}W^{+}$. And jets in the final state could be more powerful to trigger a signal at the electron positron collider than the hadron collider. So we expect that this is the best channel for probing the doubly charged leptons.

The $e^{-}+4j$ final state signature suffers some backgrounds from two standard model processes: $e^-\gamma\rightarrow e^-+4j$ and $e^-e^+\rightarrow e^-+4j+\missE$. The leading backgrounds for $e^-\gamma$ collision are final states $e^-\gamma$ and $e^-Z$ with the 2 jets decayed by the photon($Z$) and 2 jets coming from radiation. For $e^-e^+$ collision, the main backgrounds are from $q\bar{q}$ and $W^+W^-$ with subsequent decays. Other backgrounds are non-resonant which are easy to be reduced. Another possible source of reducible backgrounds is either a misidentified electron or jets. Taking into account the tight identification criteria, the expected rate of fake particles is very low \cite{ILCTDR}. The new generation detectors at the ILC and CLIC are expected to perform at least at the same level. As a consequence backgrounds from misidentification become negligible starting from the earliest selection stage.

The following basic selection cuts are applied to hadronic channel of the signal and background events:
\begin{eqnarray}
  &p_T(e)>50 {\rm ~GeV}, ~~~p_T(j)>20 {\rm ~GeV},\nonumber\\
  &\missE < 15 {\rm ~GeV}, ~~~ \Delta R_{jj,j\ell}>0.5, ~~~|\eta|\!\leq\!2.5,
\end{eqnarray}
where $p_T$ denotes the transverse momentum, and $\missE$ is the missing transverse momentum from the invisible neutrino in the final state. The electron which has the largest $p_T$ is used as a trigger for new physics searches. As there is no neutrino in the signal but there are neutrinos in the SM background, we apply a veto cut $\missE < 15$ GeV to reduce the $\ell^-4j+\missE$ events from $e^-e^+$ collisions. $\Delta R_{ij}$ is defined as $\Delta R_{ij}=\sqrt{(\Delta \eta_{i,j})^2+(\Delta \phi_{i,j})^2}$, where $\Delta\eta$ is the rapidity gap and $\Delta\phi$ is the azimuthal angle gap between the particle pair.  These base cards are typically adopted to reproduce a general purpose detector geometrical acceptance and minimal requirements to detect charged leptons. After the basic cuts, we further employ optimized kinematical cuts according to the kinematical differences between the signal and background.

\begin{figure}[!htb]
\begin{center}
\includegraphics [scale=0.38] {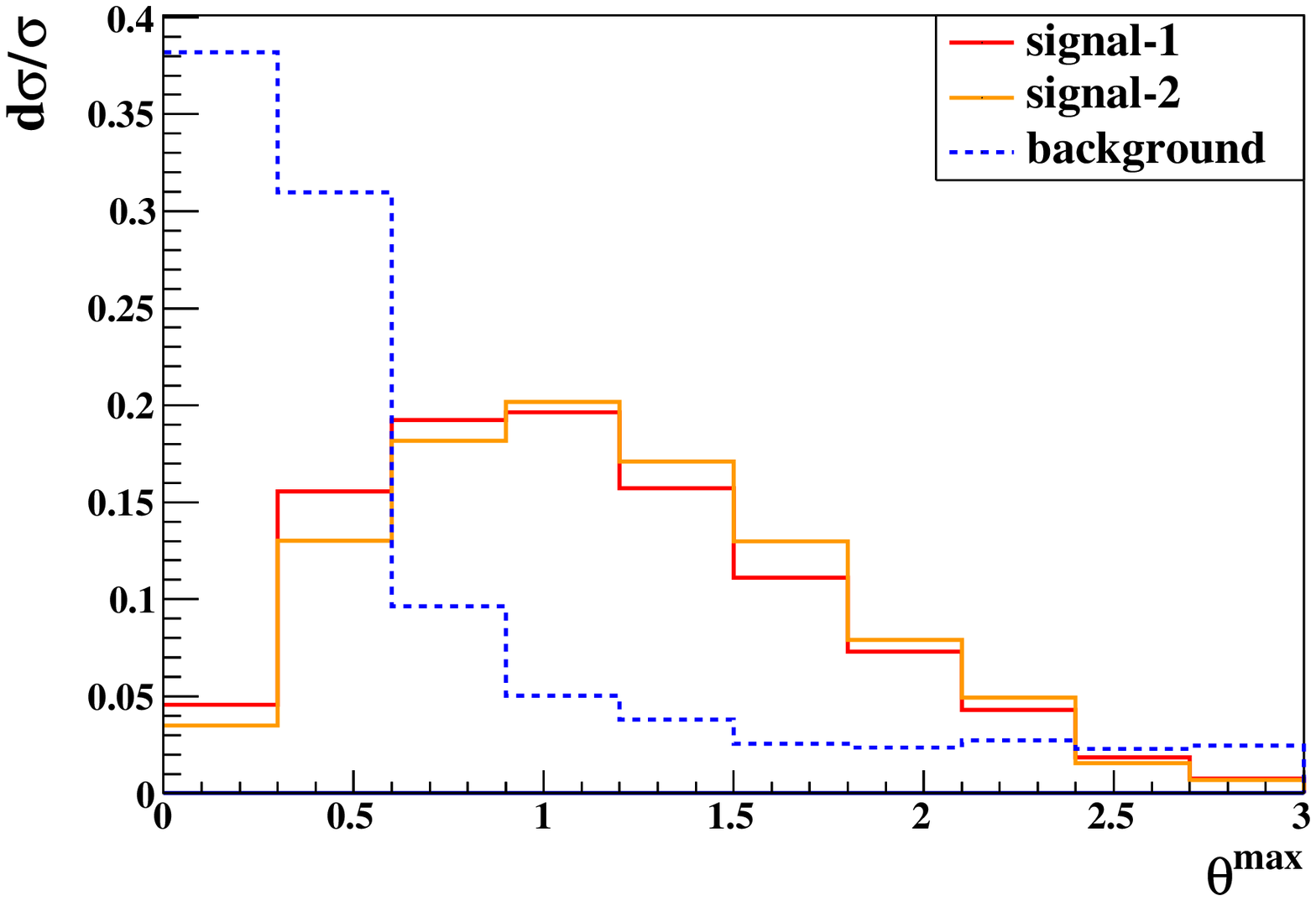}
\includegraphics [scale=0.38] {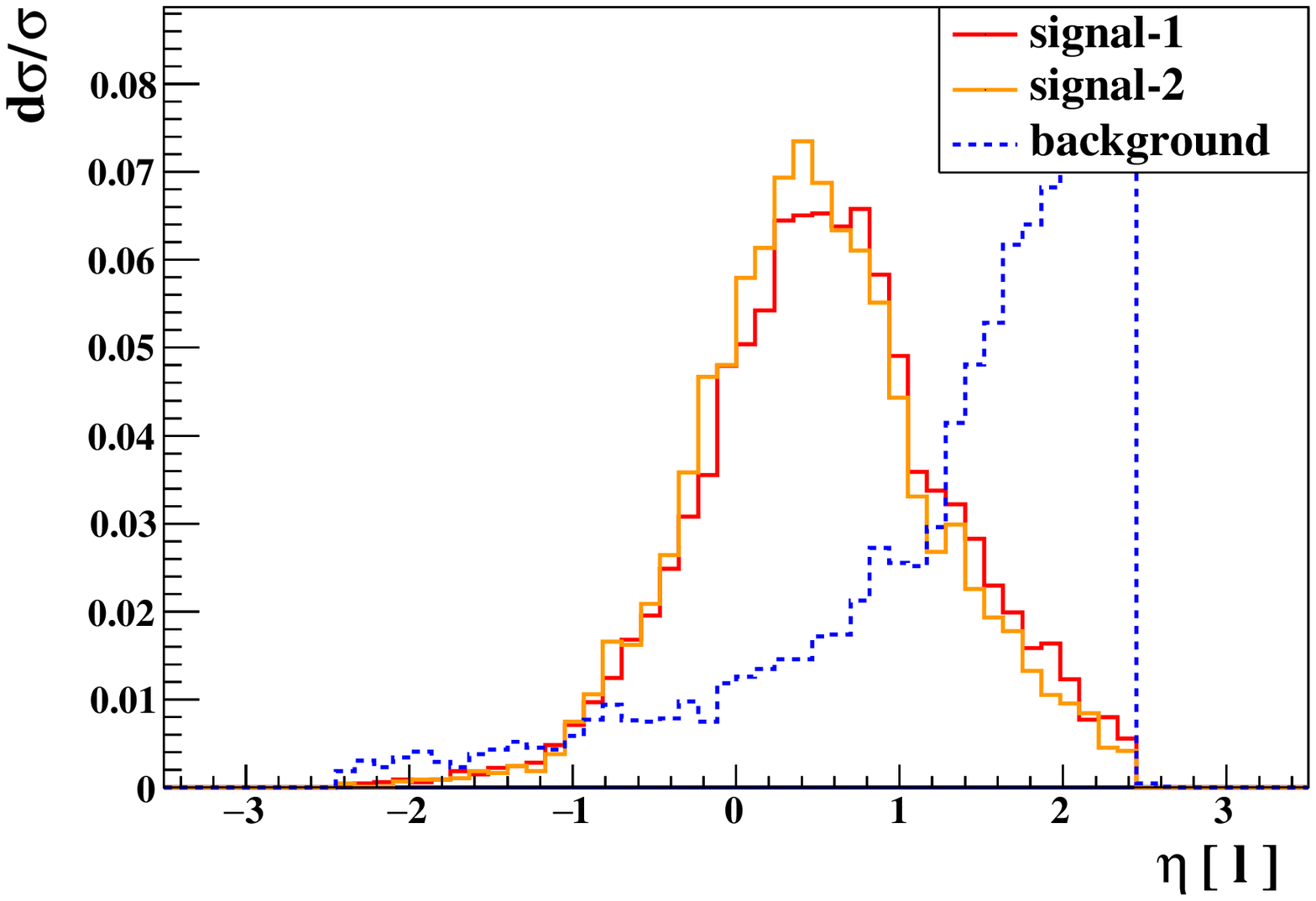}\\
\includegraphics [scale=0.38] {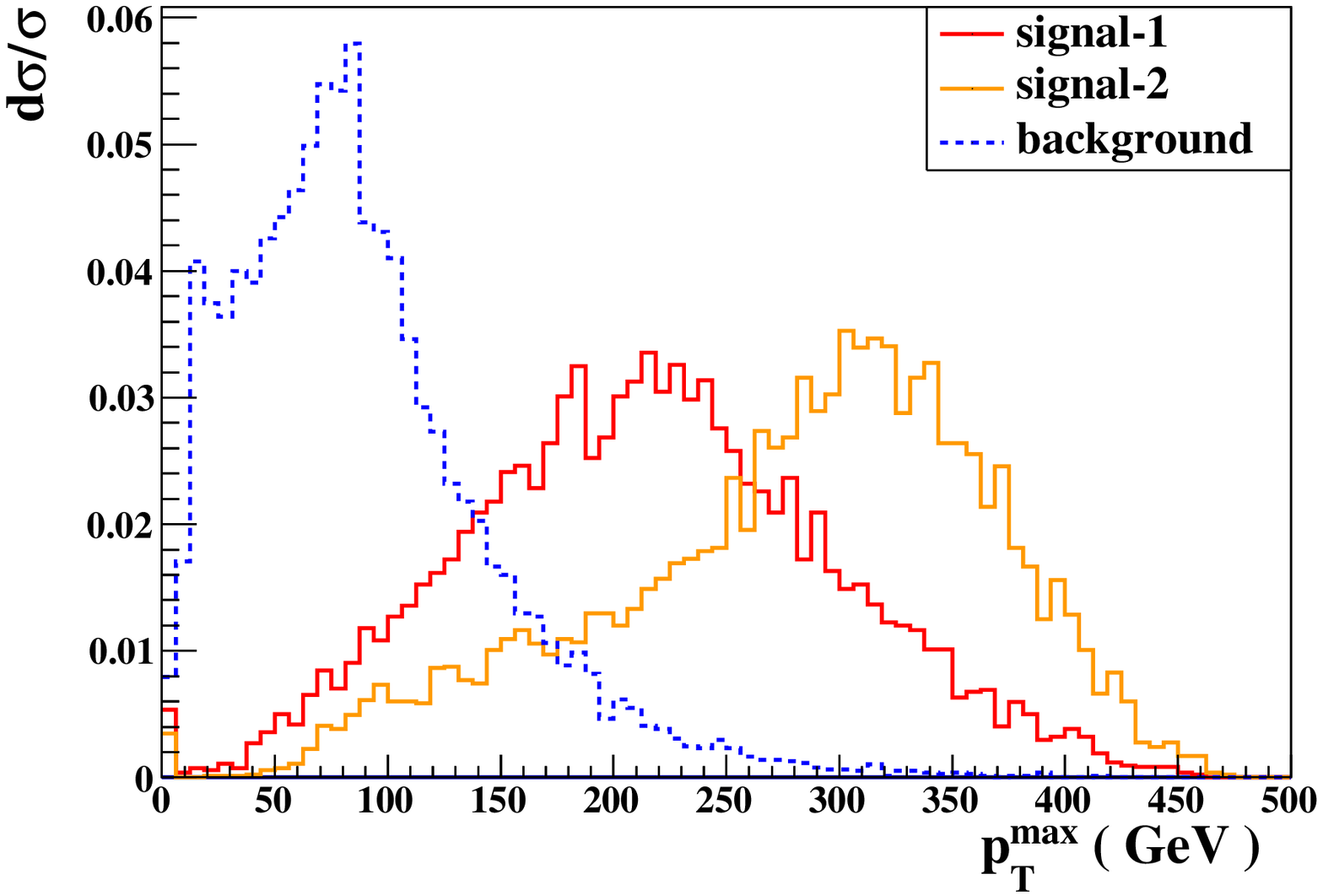}
\includegraphics [scale=0.38] {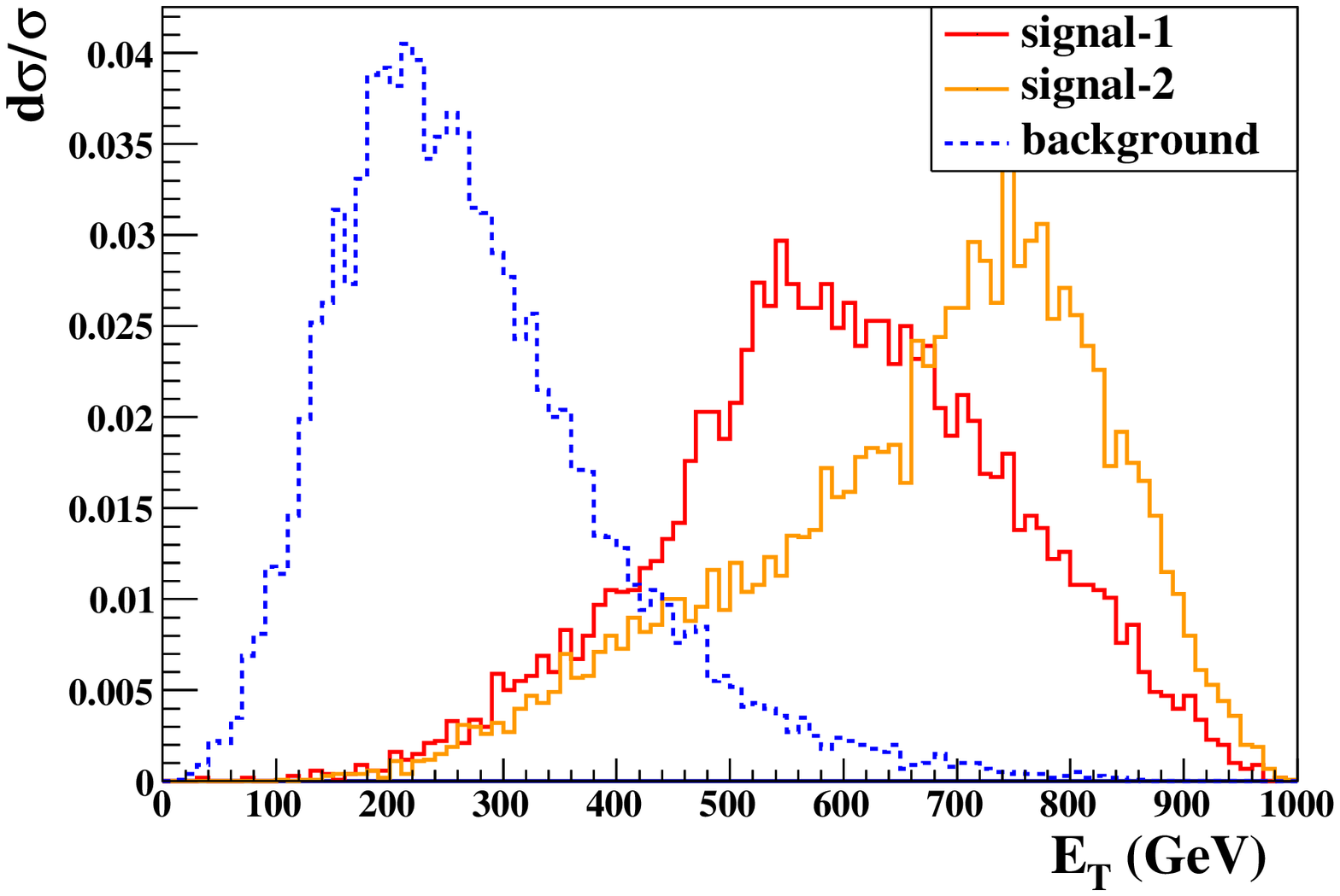}
\caption{\label{ma5:1l} Some useful kinematic observables for the hadronic channel which can separate the signal from background events are shown. Two typical cases of $M_L=500$ and 700 GeV with $\Lambda=5$ TeV are taken by the signal-1 and signal-2. No cuts are implemented to obtain these distributions.
}
\end{center}
\end{figure}

Generally speaking, the invariant mass is an appropriate kinematic variable to discriminate the signal from background. However, the mass of doubly charged lepton is one of variables of the model. A fixed mass window is unfit for discussing the statistical significance in a high mass region. Considering this process, the angular distribution of $e^-$ from $L^{--}$ decay products in the lab frame is easily reconstructed. At the same time, this is not the case for the energy distribution. As a result, the following kinematic variables are exploited to develop additional cuts: the angle $\theta$ between the electron momentum and beam axis, the pseudorapidity $\eta$ and transverse momentum $p_T$ of the electron. We also apply an important global observable, the total transverse energy $E_T$, which is related to the visible objects.

In Fig. \ref{ma5:1l}, we display the normalized distribution of these observables for some particular choices of the model parameters ($M_L=500$ and 700 GeV with $\Lambda=5$ TeV) by MadAnalysis 5 \cite{ma5}. We can see that the signal is well distinguished from the corresponding background by $\theta(e)$. The momentum of electron in the SM background is mostly along the axial direction which is different from the signal. The pseudorapidity distribution is quite convergent for the signal, whereas it is more divergent for the SM background. Due to the relatively large mass of doubly charged leptons, their decay products are largely boosted. The transverse momentum of electrons in the signal should be much higher than ones which is given by the mass of the electroweak gauge bosons in the SM background, because the momentum initially is zero on the transverse plane. Therefore we expect that $p_T(e)$ and $E_T$ distributions peake to higher values in the signal compared to background. Just as expected, the distributions show that the electron $p_T$ spectrum peaks at around half of the doubly charged lepton mass while the electrons in the SM background tend to be soft. Considering the kinematics, we impose the following improved cuts:
\begin{eqnarray}
  &p_t(e) > 140 {\rm ~GeV}, ~~~\theta(e) > 0.3 ,\nonumber\\
  & E_T > 400 {\rm ~GeV}, ~~~\eta(e) < 1.6.
\end{eqnarray}

\begin{table}[!t]
\caption{\label{t1}The number of events for both signal and background for $M_L=500$ (700) GeV at $\sqrt{s}=1$ TeV ILC with $\mathcal{L}=100$ fb$^{-1}$.  The event selection efficiency has been optimized with respect to the signal.}
\begin{tabular}{c|c|c|c}
\hline
\hline
\multicolumn{4}{c}{ILC - $\sqrt{s}= 1 $ TeV}\\
\hline
~~~Events~~~ & ~~~Signal~~~ & ~~~Bkg~~~  & ~~~$S/\sqrt{S+B}$~~~\\
\hline
No cut & 113 (53.4)  & 198  &  6.41 (3.37) \\
\hline
Basic cuts & 87.14 (40.28) & 102.93  & 6.32 (3.366) \\
\hline
$\theta(e)>0.3$, $\eta(e)<1.8$ & 69.37 (33.36) & 27.19  & 7.059 (4.287)\\
\hline
$p_t(e)>140$ GeV & 65.31 (32.87) & 10.54 & 7.50 (4.989)\\
\hline
$E_T > 400 $GeV & 63.78 (32.52) & 7.25 & 7.57 (5.156)\\
\hline
\hline
\end{tabular}
\end{table}

We calculate the expected statistical significance $S/{\sqrt{(S+B)}}$ (SS) for the luminosity of 100 fb$^{-1}$ at the ILC, where $S$ and $B$ denote the number of the signal and background events, respectively. The results for each cut are summarized in Table \ref{t1}. The former data in all columns and those in parentheses correspond to the results for the doubly charged lepton mass $M_L=500$ and $M_L=700$ GeV, respectively. It is obvious that the sets of cuts are powerful in signal event selection. The significance can reach 7.57 (5.156) for $M_L=500$ (700) GeV and $\Lambda=5$ TeV at the 1 TeV ILC with an integrated luminosity of 100 fb$^{-1}$. For $M_L=500$ (700) GeV, the $5\sigma$ significance requires 43.63(94.04) fb$^{-1}$. Obviously, this channel is the most promising one to probe doubly charged leptons via $L^{--}W^+$ production in $e^+e^-$ colliders. We further consider the hadronic signal at the CLIC, where the cross section of background is much smaller than that of the signal. Thus we only apply the basic cuts on the signal and background. For $M_L=500$ (700) GeV and $\Lambda=5$ TeV, the statistical significance can reach as high as 27.762(30.849).

\begin{figure}
\begin{center}
\includegraphics [scale=0.8] {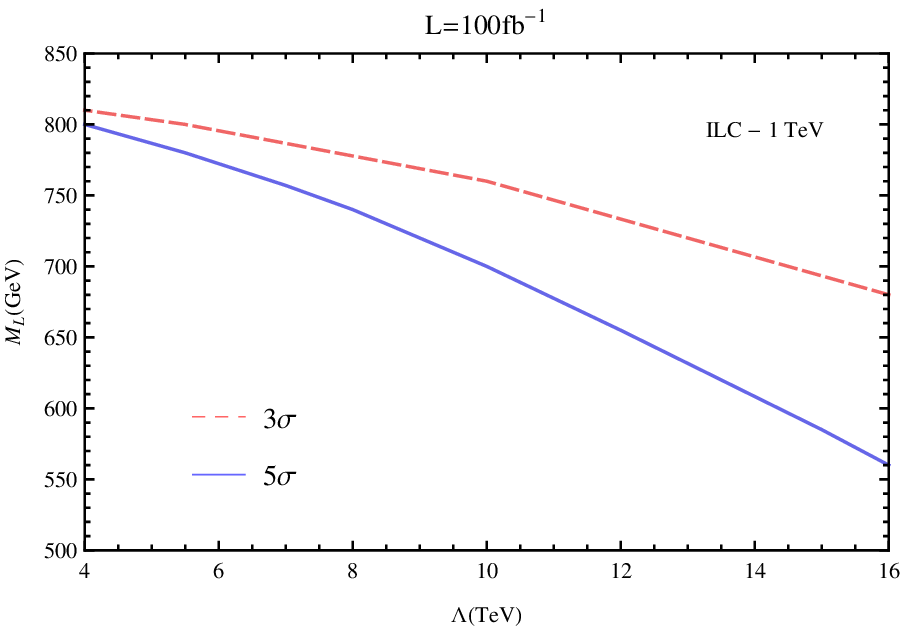}~~
\includegraphics [scale=0.8] {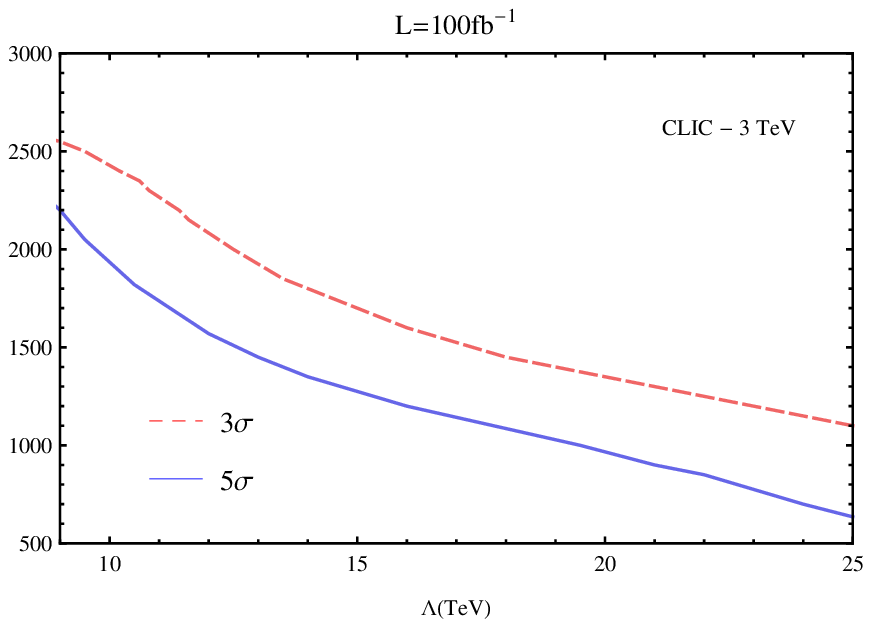}\\
\includegraphics [scale=0.8] {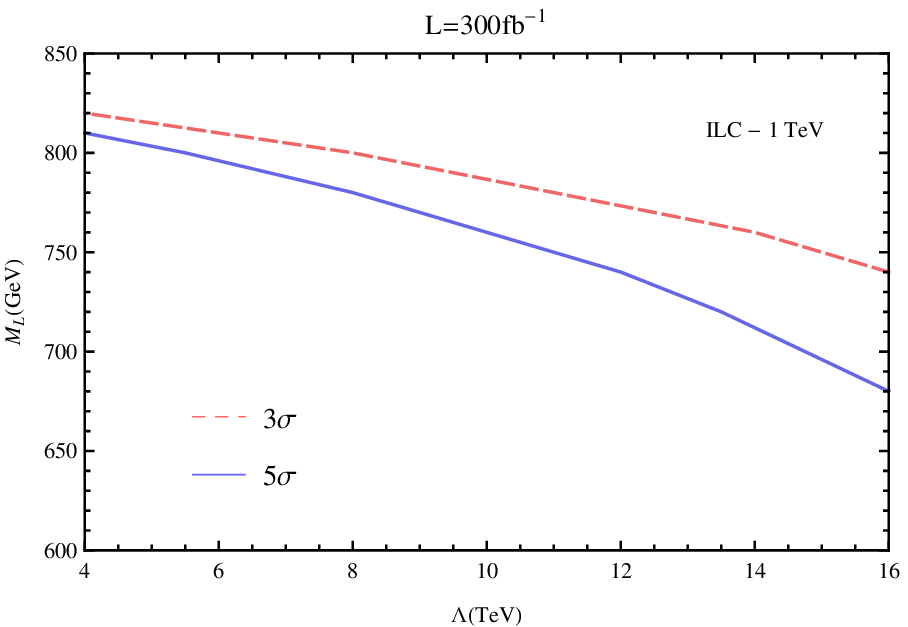}~~
\includegraphics [scale=0.8] {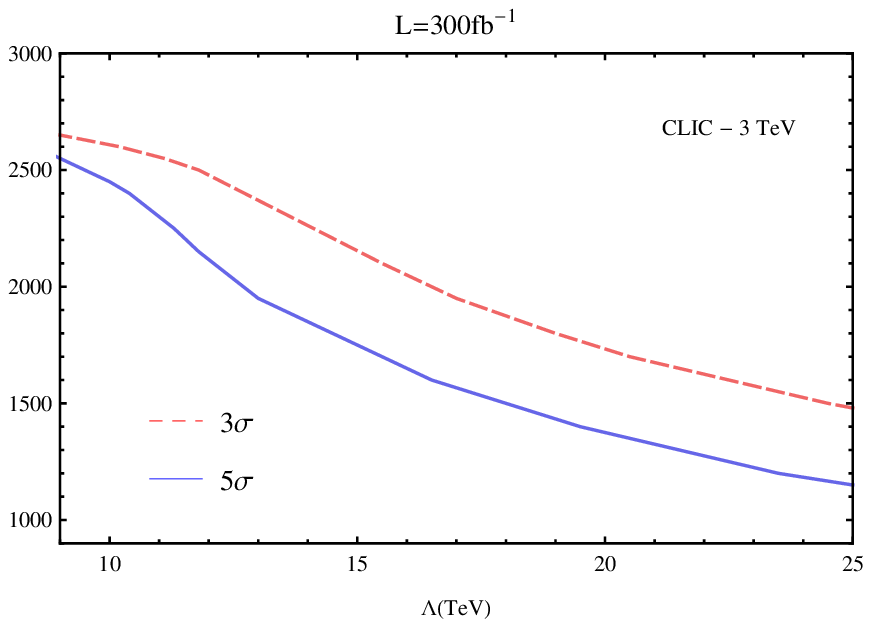}\\
\includegraphics [scale=0.8] {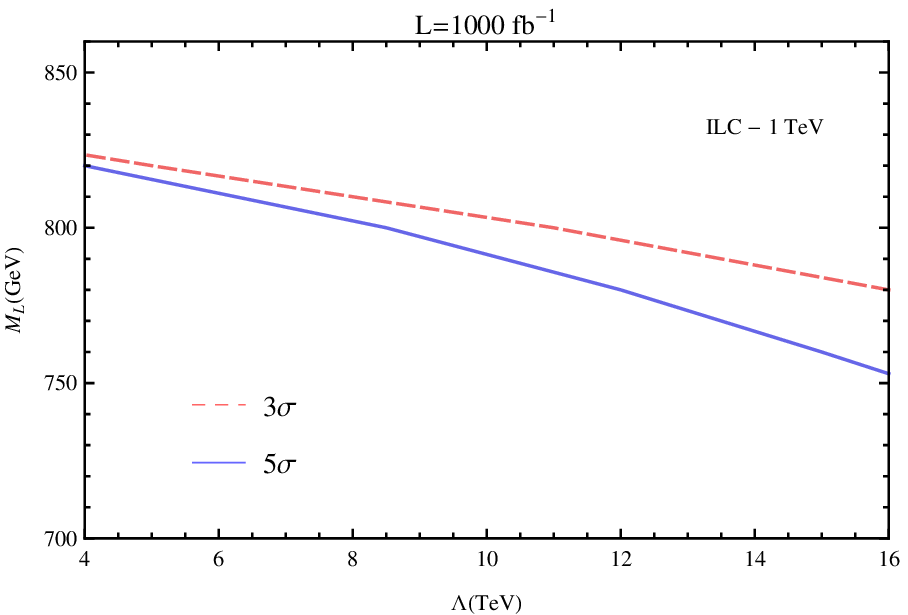}~~
\includegraphics [scale=0.8] {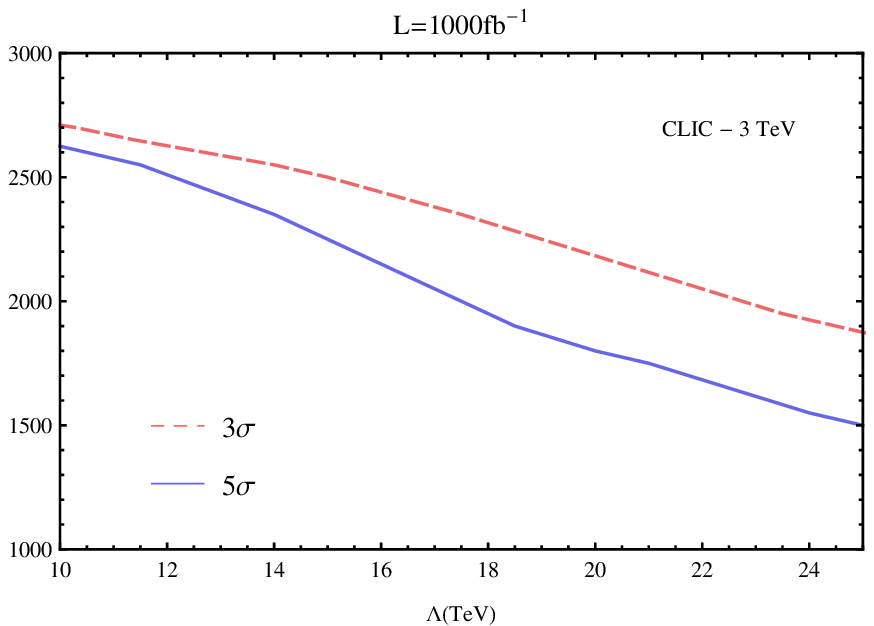}\\
\includegraphics [scale=0.8] {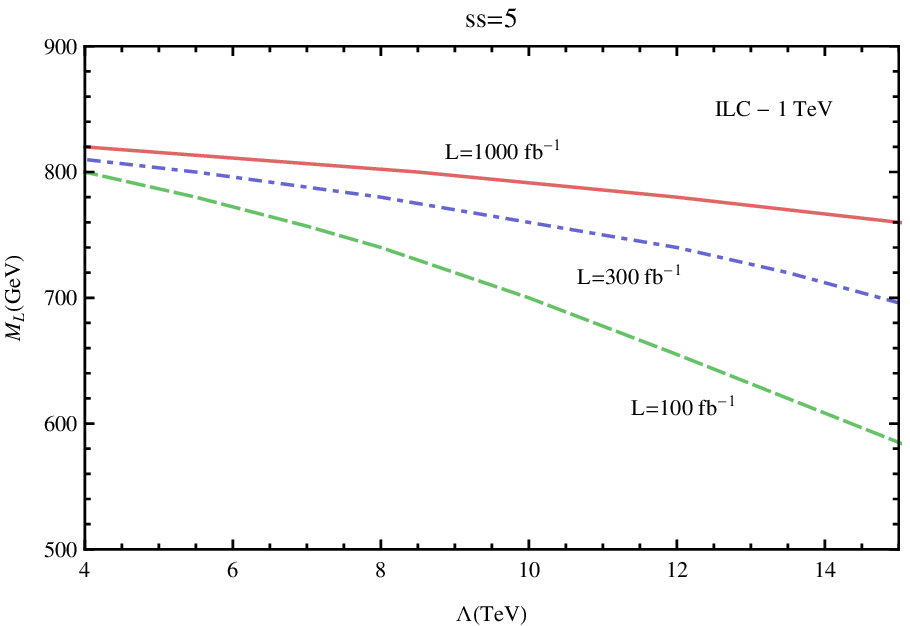}~~
\includegraphics [scale=0.8] {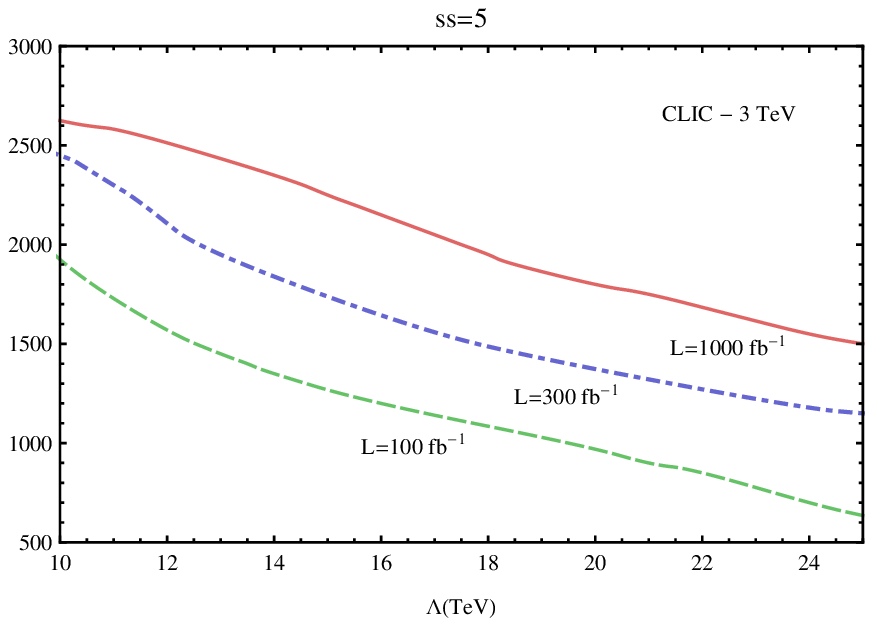}
\caption{\label{ss:1l} 3$\sigma$ and $5\sigma$ contour plots for the single lepton signal in the $\Lambda-M_L$ plane at the 1 TeV ILC and 3 TeV CLIC with three values of the integrated luminosity $L=$100, 300, 1000 fb$^{-1}$.}
\end{center}
\end{figure}

Performing the scanning over the parameter space of $M_L$ and $\Lambda$, we obtain the experimental evidence region (SS $\geq3$) and experimental discovery region (SS $\geq5$). We consider three different cases for the nominal luminosity, $L=$100, 300, 1000 fb$^{-1}$, and two energies of the colliding electrons, $\sqrt{s}=1$ TeV and $\sqrt{s}=3$ TeV. In Fig. \ref{ss:1l}, we plot the parameter space regions accessible by the ILC and CLIC at $3\sigma$ and $5\sigma$ levels with three values of the luminosity given above.
The $3\sigma$ exclusion regions at the ILC and CLIC are compared with the bounds provided by the previous work \cite{1411.6556} and current LHC analyses \cite{cms1,atlas1,cms3,cms4,atlas4}. The authors of Ref. \cite{1411.6556} analyze the potential of the linear collider to search for doubly charged leptons via $e^-e^-\rightarrow E^{--}\nu_e$ process, showing that the ILC and CLIC will be able to set direct constraints on these new states in the parameter space at different luminosities.  We find a good agreement comparing our results with this work. We also notice that our results turns out that the expected ILC and CLIC exclusion bounds on $\Lambda$ the are much better than the current LHC bound, perhaps investigating possible effects due to the available polarization of the beams that could still improve the bounds of Fig. \ref{ss:1l}.
Our CLIC bound on $\Lambda$ for $M_L\approx 2$ TeV ($\Lambda >$ 16-17 TeV) at the 3-sigma level is highly competitive with the predictions at 14 TeV LHC \cite{cms3} where for $M_L\approx 2$ TeV the lower bound on the compositeness scale is supposed to be $\Lambda >$ 11.6 TeV with an integrated luminosity of 300 fb$^{-1}$.
The reasons should be attributed to much cleaner backgrounds and single production of doubly charged lepton, which are the advantage in terms of sensitivity for the electron-photon mode.
Actually, the actual bounds of excited leptons with ordinary isospin, properly speaking, do not apply to the extended isospin multiplets with double charges. Nevertheless, we believe that the bounds on doubly charged leptons should not be so different with current experimental analyses.

Based on the above results, the possible signature of doubly charged leptons is limited in the lower mass range and could be detected only with high luminosity at the ILC. On the other side, the CLIC collider could offer better detection capabilities than existing colliders within the high mass range for searching heavy doubly charged leptons.

\subsection{Semi-leptonic Channel}

The bilepton candidate events are selected from the the production processes of
\begin{eqnarray}
e^{-}\gamma\rightarrow L^{--}W^{+}\rightarrow e^{-}\ell^{-}\bar{\nu}jj(e^{-}\ell^{+}\nu jj),
\end{eqnarray}
in which one of the $W$ bosons decays leptonically and the other one decays hadronically. Consequently, two kinds of signals are provided: opposite sign charged leptons or same sign charged ones, depending on which $W$ decays leptonically.
The backgrounds of same sign charged bi-lepton are the final states $e^-\gamma(Z)$ and $\nu_{e}W^-$ with subsequent decays which are described by a total of 426 Feynman diagrams. For opposite sign charged final state, backgrounds for $e^{-}\ell^{+}+2j+\missE$ signature are produced by $e^-\gamma$, $e^+e^-$ or $\gamma\gamma$ collisions and larger than same sign charged one. Thus we will focus on the bi-lepton signature with a same sign charge which is characterized by a low SM background.

\begin{figure}[!htb]
\begin{center}
\includegraphics [scale=0.38] {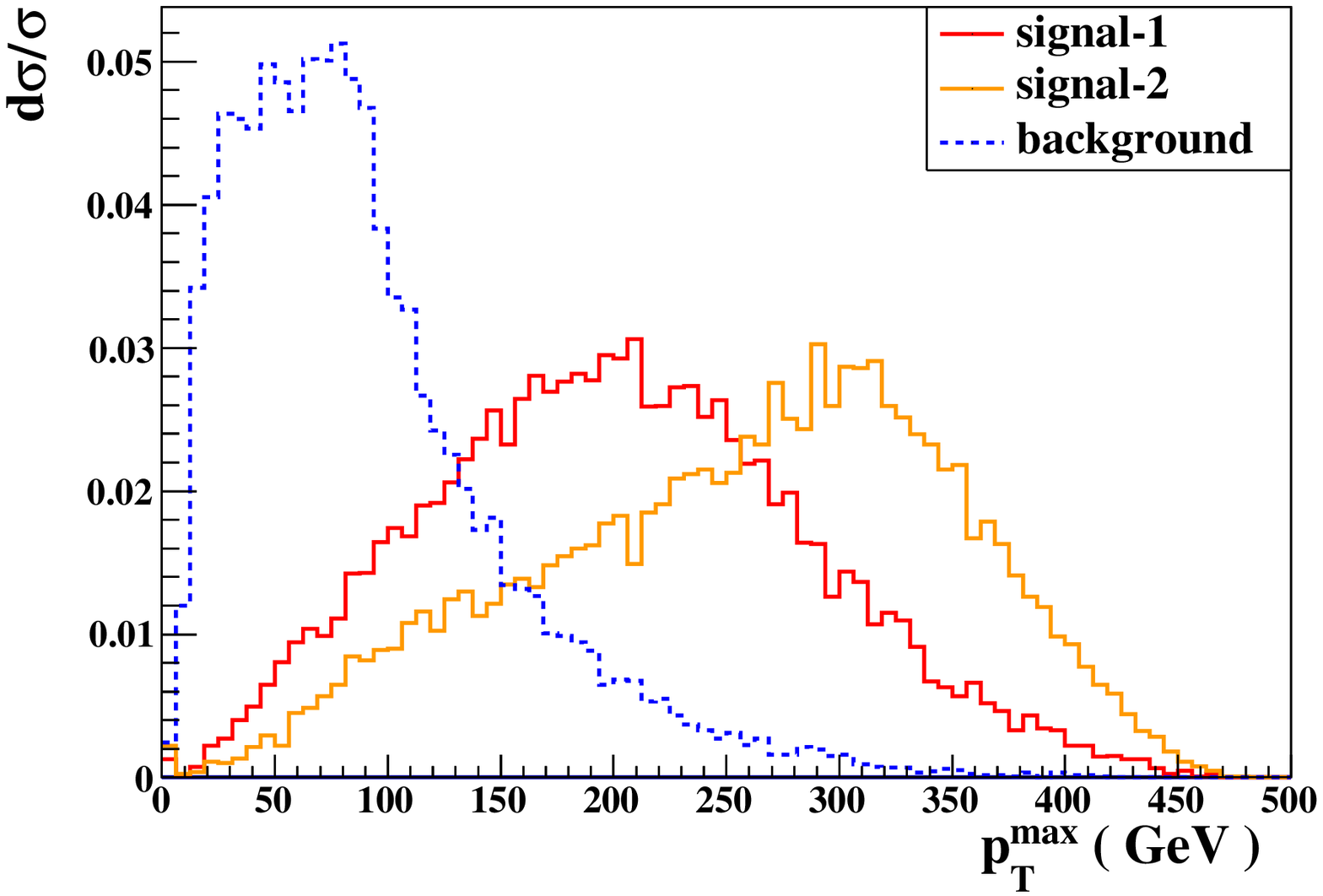}
\includegraphics [scale=0.38] {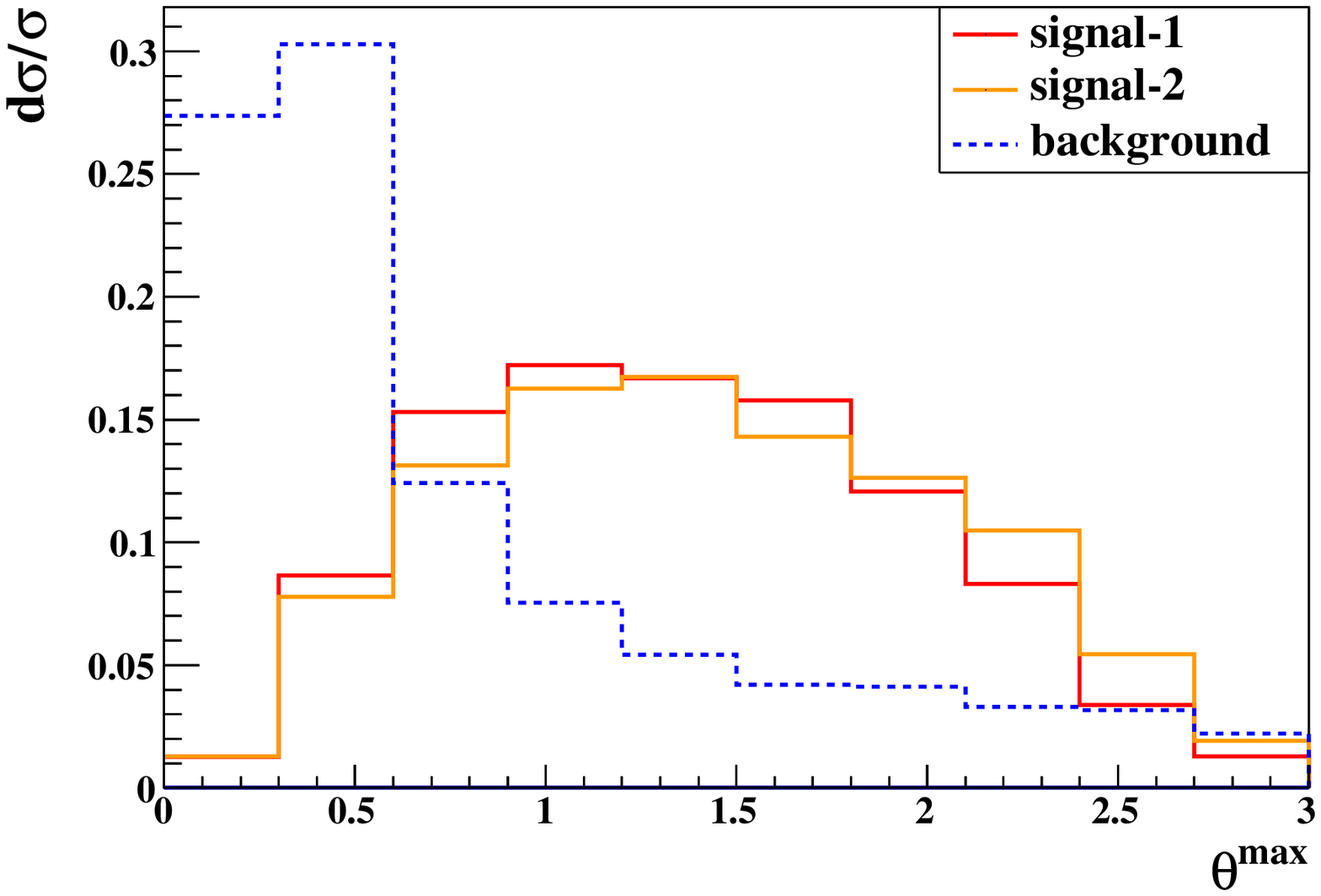}\\
\includegraphics [scale=0.38] {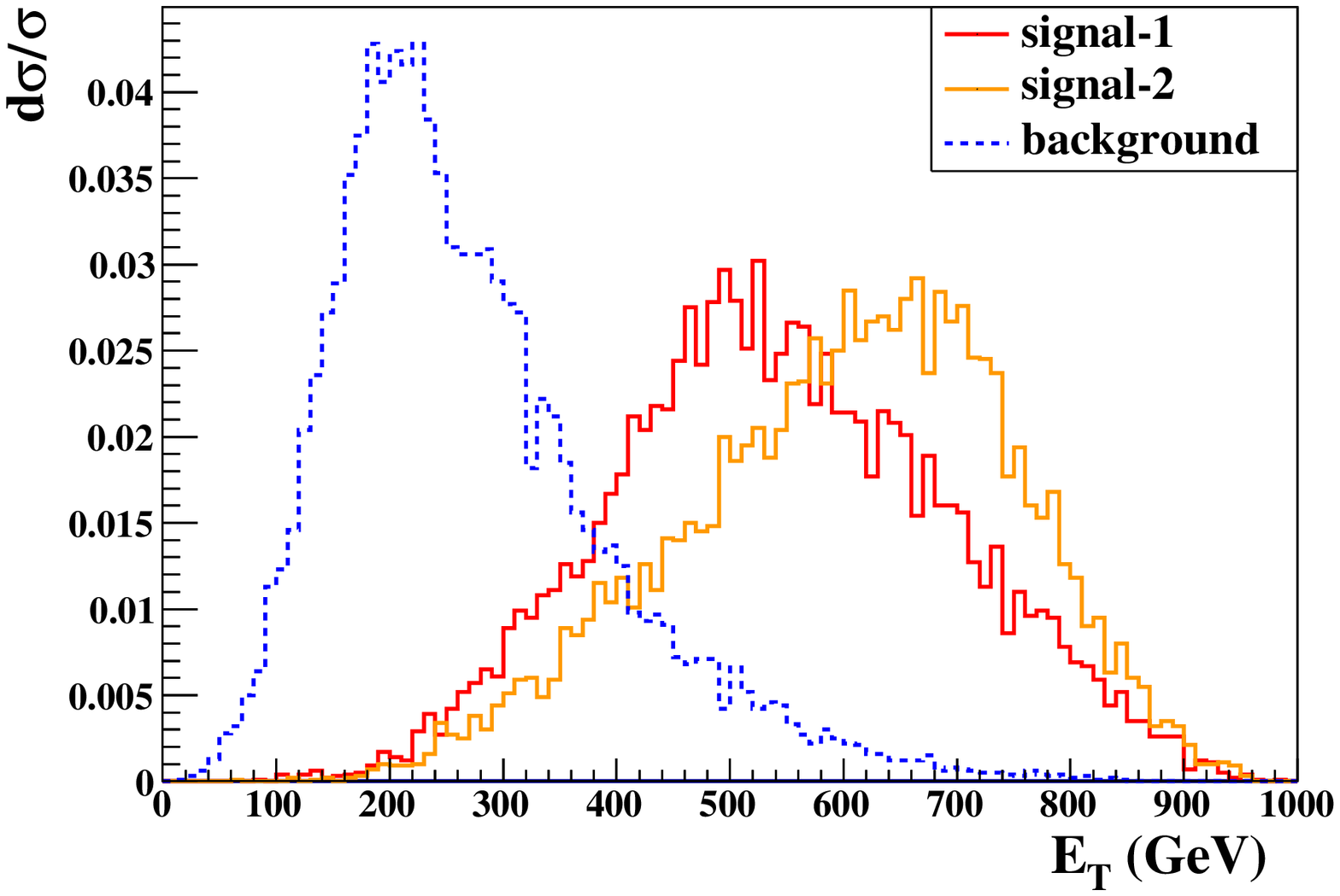}
\includegraphics [scale=0.38] {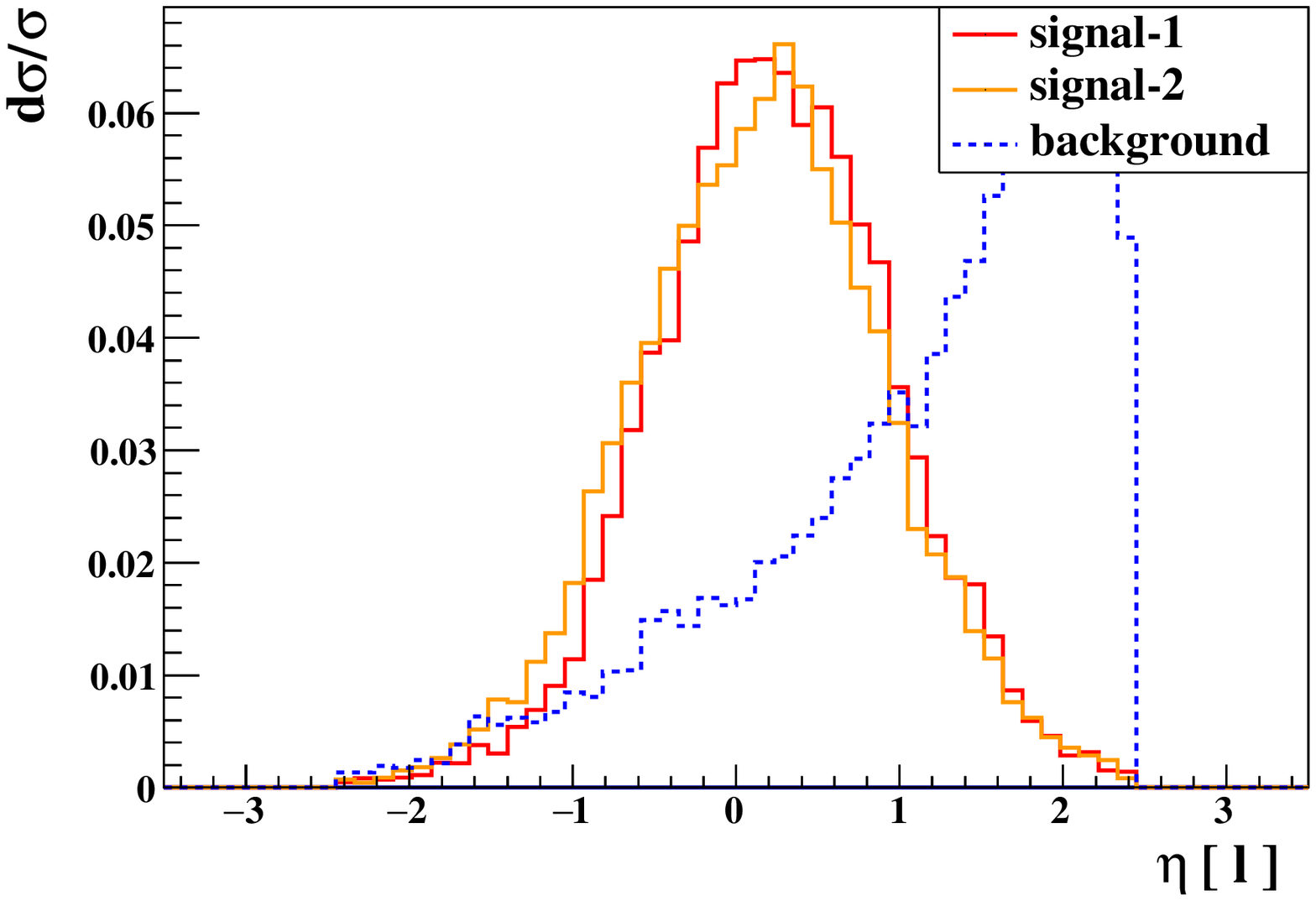}\\
\includegraphics [scale=0.38] {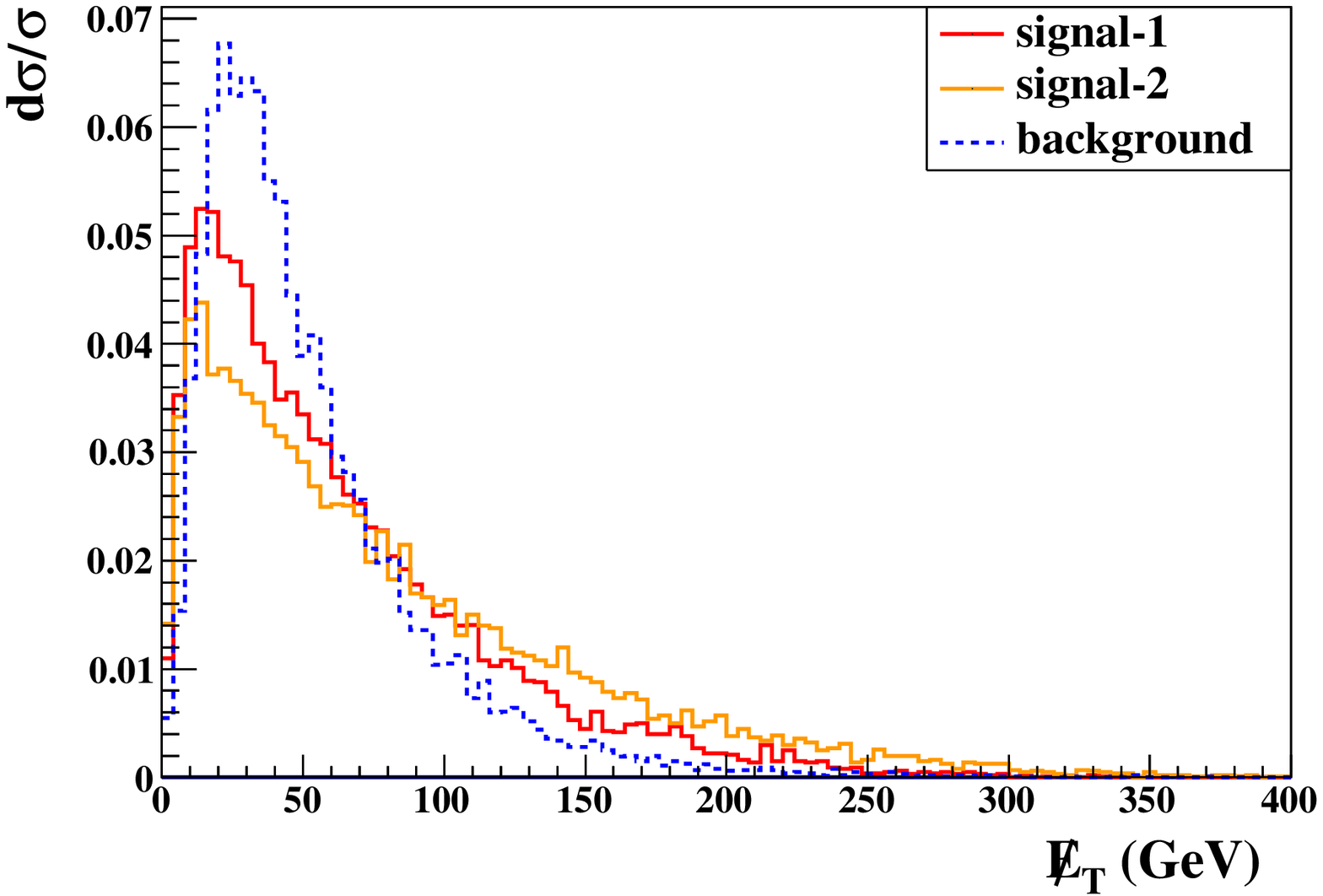}
\includegraphics [scale=0.38] {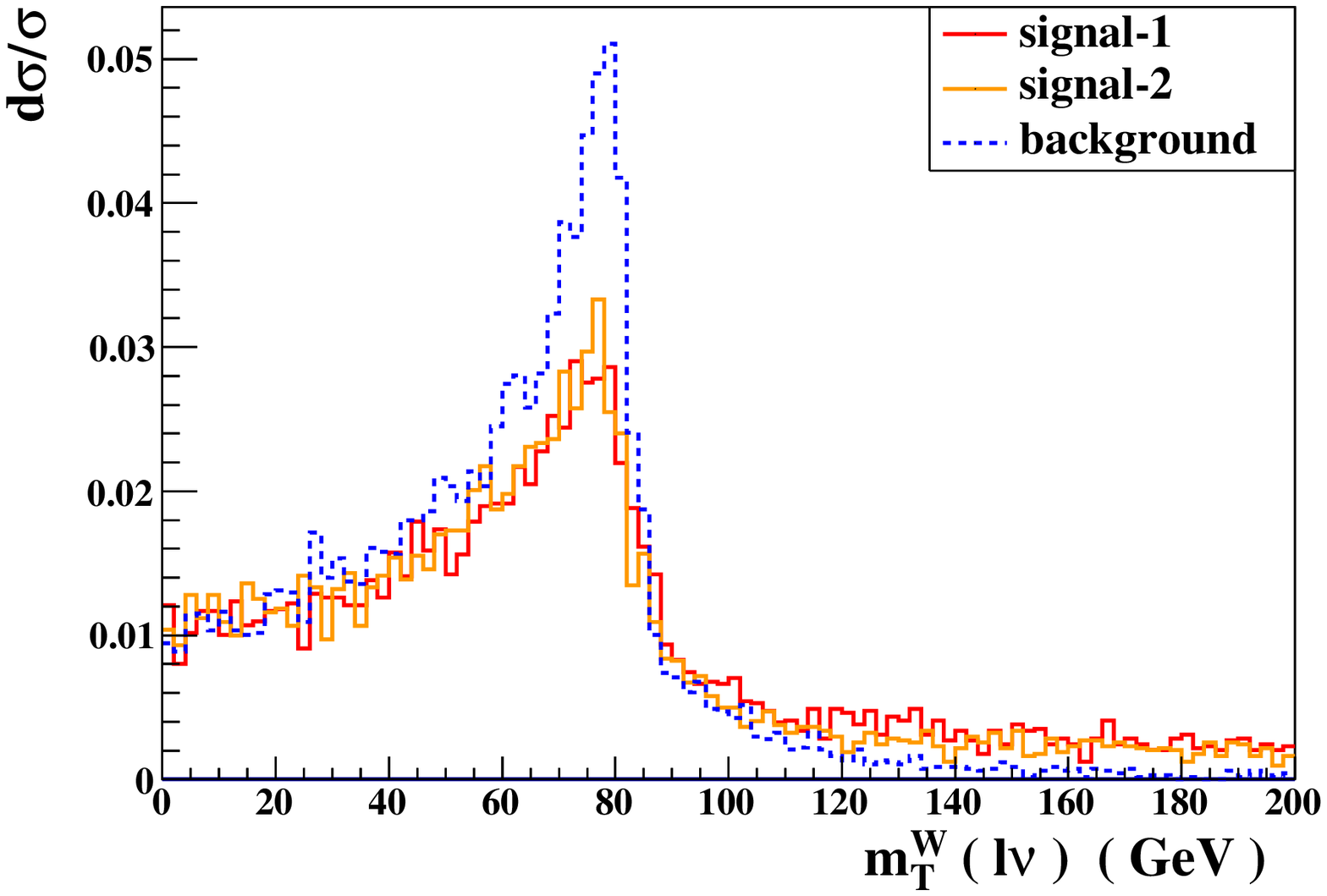}
\caption{\label{ma5:2l} The kinematic variables for the same sign charged bi-lepton signal and the SM background with $M_L=500$ (signal-1) and 700 GeV (siganl-2) are shown.}
\end{center}
\end{figure}

The basic cuts on final state transverse momentum, rapidity and separations are applied to bi-lepton signal and background events:
\begin{eqnarray}
  &p_t(\ell)>15 {\rm ~GeV}, ~~~p_t(\ell_1)>50 {\rm ~GeV}, ~~~p_t(j)>20 {\rm ~GeV} ,\nonumber\\
  & \missE>15 {\rm ~GeV}, ~~~|\eta|\!\leq\!2.5, ~~~\Delta R_{jj,\ell\ell,j\ell}>0.5.
\end{eqnarray}
Then, we give a kinematic characterization of the signal and the corresponding background in terms of several variables as discussed above in Fig. \ref{ma5:2l}. In view of characteristic of the semi-leptonic channel, we examine the normalized missing transverse energy $\missE$ related to the invisible objects, and reconstruct the mass of the $W$ boson which decays leptonically. From Fig. \ref{ma5:2l}, these two new cuts will not be very effective in further suppressing the SM background, but the electrons in the signal are more energetic than those of the irreducible background, and the cuts on these variables could greatly suppress background processes except for the $m_T^W$. So we impose the kinematical cuts as follows:
\begin{eqnarray}
&~~~~~E_T > 350 {\rm GeV},~~~p_T^{\rm max} > 140{\rm GeV},\nonumber\\
&~\eta^{\rm max} <1.8, ~~~~~~~~\theta^{\rm max} > 0.5.
\end{eqnarray}

After taking into account all these cuts, we can further suppress background and gain in the significance up to $\sim3\sigma$, as represented in the Table \ref{t2}. We also calculate the needed integrated luminosity for a discovery by the semi-leptonic channel at the 1 TeV ILC (see Fig. \ref{L-M}, left panel). The results show that the potential to observe doubly charged leptons can not match the former optimal hadronic channel.

\begin{table}[!htb]
\caption{\label{t2}The event numbers of the bi-lepton signal and the corresponding background for $M_L=500$ (700) GeV and $\Lambda=5$ TeV at $\sqrt{s}=1$ TeV with $\mathcal{L}=100$ fb$^{-1}$. }
\begin{tabular}{c|c|c|c}
\hline
\hline
\multicolumn{4}{c}{ILC - $\sqrt{s}= 1 $ TeV}\\
\hline
~~~Events~~~ & ~~~Signal~~~ & ~~~Bkg~~~  & ~~~$S/\sqrt{S+B}$~~~\\
\hline
No cut & 50.7 (23.4)  & 753  &  1.79 (0.84)  \\
\hline
Basic cuts & 7.45 (5.06)  & 225.4 & 0.488 (0.33)   \\
\hline
$\eta^{\rm max} < 1.4$, $\theta^{\rm max} > 0.5$ & 7.45 (5.02)  & 194.1 & 0.525 (0.357)  \\
\hline
$p_T^{\rm max}>140$ GeV & 6.56 (4.76)  & 45.03 & 0.913 (0.675)  \\
\hline
$E_T > 350 {\rm GeV}$  & 6.23 (4.66)  & 28.64 & 1.055 (0.807)  \\
\hline
\hline
\end{tabular}
\end{table}

\subsection{Pure leptonic Channel}
Although the leptonic channel of $L^{--}W^+$ production provides the minimal cross section among all channels, this channel also provides special final state that include three leptons and large missing transverse energy $\missE$,
\begin{eqnarray}
e^{-} \gamma \rightarrow L^{--} W^{+} \rightarrow e^{-}\ell^{-}\ell^{+}\nu\bar{\nu}.
\end{eqnarray}

For signal signature $3\ell+\missE$, the corresponding background is described by a total of 972 Feynman diagrams which come from the $e^-\gamma\rightarrow e^-\gamma(Z)$ and $e^-\gamma\rightarrow\nu_{e}W^-$ with subsequent decays.
After considering the basic cuts which are the same as those in the semi-leptonic channel (without any cuts on jets), we further employ optimized kinematical cuts as follows:
\begin{eqnarray}
&~~E_T > 350 {\rm GeV},~~~p_T^{\rm max} > 140{\rm GeV},\nonumber\\
&\theta^{\rm max} >0.5 , ~~~~~\eta^{\rm max} < 1.5.
\end{eqnarray}

\begin{figure}[!htb]
\begin{center}
\includegraphics [scale=.83] {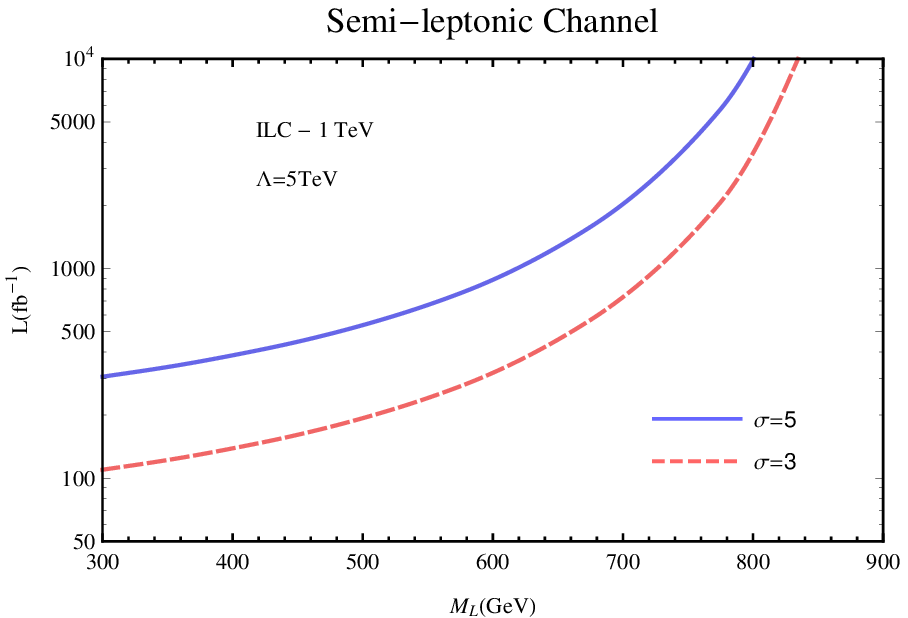}
\includegraphics [scale=.83] {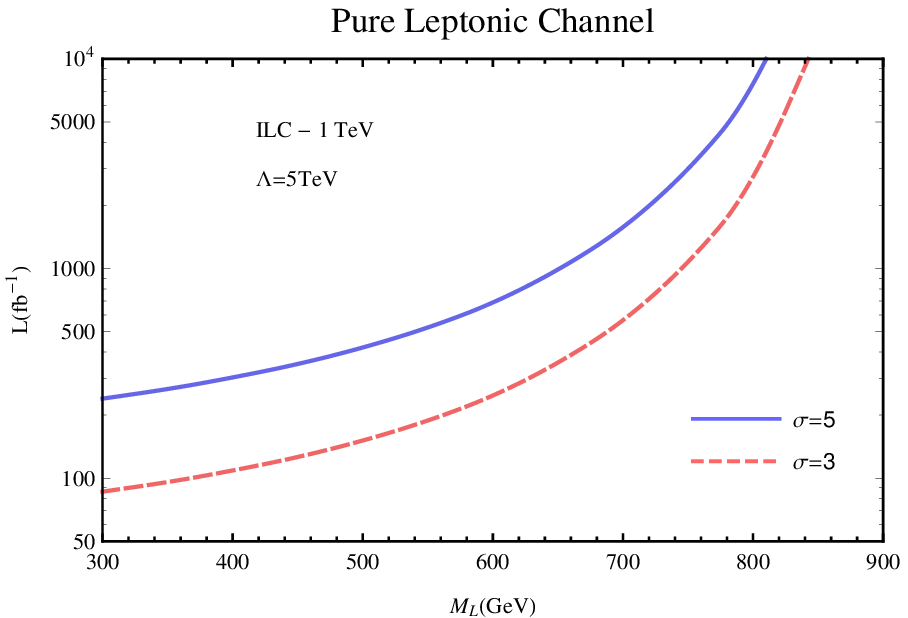}
\caption{\label{L-M}The needed luminosity to observe different masses of the doubly charged lepton via the semi-leptonic channel (left) and pure leptonic channel (right) for $3\sigma$ and $5\sigma$ statistical significances at the 1 TeV ILC.} \label{}
\end{center}
\end{figure}

We calculate the statistical significance for the luminosity of 100 fb$^{-1}$ at the ILC. The results revealed that it is challenging to discover the tri-lepton signature for the pure leptonic channel at the ILC for $M_L=500$ (700) GeV, with $2\sigma$ ($1\sigma$) level statistical significance. Thus, we do not show the relevant numerical results.
The needed integrated luminosity for observing doubly charged leptons with mass $M_L$ for $3\ell+\missE$ signal is plotted in Fig. \ref{L-M} (right). For $M_L>800$ GeV, detecting this signal at 5$\sigma$ requires an integrated luminosity larger than $10^4$, which outreaches the designed luminosity. The needed luminosity for the ILC to observe the doubly charged leptons via this channel is rather large, without any advantage compared to the hadronic channel.

\section{Conclusions }
The existence of heavy doubly charged leptons are predicted by higher isospin multiplets, $I_{W}=1$ and $I_{W}=3/2$, in the extended isospin model. Such new particles are expected by an assumption that the internal structure exists. One of the design purposes of linear electron positron colliders is searching for doubly charged leptons and making precise measurements.

In this paper, we investigate the single production of doubly charged leptons in forthcoming linear colliders. In particular we focus on the single production by $e^-\gamma$ collisions, a possible option for the linear colliders, that allows a wider mass parameter space to be probed than pair production by $e^+e^-$ collisions. We present the cross sections for the $e^-\gamma \rightarrow L^{--}W^{+}$ process depending on doubly charged lepton mass at 1 TeV ILC and 3 TeV CLIC.

By considering the different decay modes of the $W$ boson, we study the hadronic, semi-leptonic and pure leptonic channels in detail. In addition, we apply different basic cuts and chose several useful observables according to the kinematical differences between the signal and the relevant SM background based on the full simulation performance of the SiD. By comparing all the channels, the hadronic channel is the best one which can provide a single lepton final state to detect doubly charged leptons. Therefore, we provide the $3\sigma$ and $5\sigma$ statistical significance exclusion curves in the $\Lambda-M_L$ parameter space. Compared to the existing researches as regards single doubly charged lepton productions in future linear electron positron colliders \cite{1405.3911,1411.6556}, our results give more sensitive constraints on mass $M_L$ and compositeness scale $\Lambda$. Our results show that the ILC can only detect possible signature in the lower mass range with a high luminosity. Nevertheless, even if the ILC has not sufficient energy to produce heavy doubly charged leptons, the virtual doubly charged leptons will also contribute to indirect searchs. In comparison, the CLIC offers better detection potential to search doubly charged leptons than existing colliders in a high mass region.

\section*{Acknowledgement}

\noindent
One of us (Y. C. G.) would thank to Zhen-hua Zhao for reading the manuscript. This work was supported in part by the National Natural Science Foundation of China under Grant No. 11275088 and Grants No. 11545012, the Natural Science Foundation of the Liaoning Scientific Committee (No.2014020151).

\end{document}